\documentstyle[twocolumn,aps,epsfig]{revtex}
\begin{document}
\draft
\title{Universality for 2D Wedge Wetting}

\author{A.\ O.\ Parry, C.\ Rasc\'{o}n, A.\ J.\ Wood}
\address{Mathematics Department, Imperial College\\
180 Queen's Gate, London SW7 2BZ, United Kingdom}
\date{\today}
\maketitle

\center{\bf Phys. Rev. Lett. 83, 5535 (1999)}

\begin{abstract}
We study 2D wedge wetting using a continuum interfacial Hamiltonian
model which is solved by transfer-matrix methods. For
arbitrary binding potentials, we are able to exactly calculate the wedge
free-energy and interface height distribution function and, thus,
can completely classify all types of critical behaviour. We show
that critical filling is characterized by strongly universal 
fluctuation dominated critical exponents,
whilst complete filling is determined by the geometry rather than
fluctuation effects. Related phenomena for interface depinning from
defect lines in the bulk are also considered.
\end{abstract}

At present, experimental methods allow the shape of solid surfaces to
be controlled at a nanoscopic level \cite{Surfaces}. Fluids 
confined by such structured substrates can exhibit quite distinct adsorption
characteristics compared to that occuring for planar systems \cite{Rough}.
This includes new
types of interfacial phase transitions and critical phenomena which are not
only of fundamental interest but may well play an important role in developing
technologies such as super-repellent surfaces \cite{Repellent},
self assembly of three-dimensional structures \cite{Assembly} or
micro-fluidics \cite{Microfluidics}, among others \cite{Barrat}.
An interesting example of these phenomena which is recently attracting
new interest is the so called {\it filling} or wedge wetting
transition of a fluid adsorbed in a wedge \cite{Wedge,WedgeOld},
formed by the junction of two flat
walls tilted at angles $\pm \alpha$ to the horizontal as shown in Fig.\ 1.
Thermodynamic arguments predict that a wedge-gas interface is completely
filled by a liquid phase (at bulk liquid-gas coexistence) for temperatures
$T>T_{\alpha}$, where the filling temperature $T_{\alpha}$ is lower than
the wetting temperature $T_{w}$ of the planar ($\alpha\!=\!0$) wall
\cite{Hauge}. In fact, according to these macroscopic arguments, the
location of the filling transition phase boundary is beautifully expressed
in terms of the contact angle $\Theta_{\pi}(T)$ of the liquid drop on
the planar substrate \cite{Hauge}:
\begin{equation}
\label{one}
\Theta_{\pi}(T)=\alpha.
\end{equation}

Thus, the liquid completely wets the wedge when the contact angle is
smaller than the tilted angle $\alpha$. 
Interestingly, this macroscopic result was predicted and
confirmed experimentally \cite{Mathematics} eight years before
the seminal paper by Cahn on wetting in planar surfaces \cite{Cahn}.

Recently, the macroscopic prediction (\ref{one}) has been supported
by mean-field analysis of a model system 
which also suggest that the qualitative order of the filling transition
(first-order or continuous) follows that of the planar wetting
transition \cite{Wedge}. Thus, for planar substrates exhibiting critical
wetting transitions, the wedge offers two new examples of interfacial-like
critical phenomena in which the interface height $\ell_0$ (measured from
the bottom of the wedge) diverges as the temperature and chemical potential
are varied. Borrowing from the vocabulary used for wetting, we refer to the
wedge filling transition ocurring as $T\rightarrow T_{\alpha}^{-}$ (at bulk
coexistence) as {\it critical filling}. In contrast, by {\it complete
filling}, we refer to the divergence of $\ell_0$ for temperatures
$T>T_{\alpha}$ as the bulk chemical potential $\mu$ is increased towards
saturation $\mu_{sat}(T)$. However, apart from a few limited results
available for the corner wetting transitions (corresponding to $\alpha=\pi/4$ 
and restricted to short range forces) \cite{Exact}, there has been no discussion in
the literature of fluctuation effects, scaling regimes and universality
classes for such filling transitions and how these compare with the rich
phenomenology known for wetting \cite{Lipowsky}. 
 With this aim in mind, we have studied an
effective interfacial Hamiltonian model of filling in (bulk) dimension
$d=2$ and derived exact elegant results for various quantities of interest
such as the excess wedge free-energy and the probability distribution for
the interface height. The formal analysis can be carried through for
{\it arbitrary} choices of binding potential ({\it i.e.}, all range of forces)
allowing a complete classification of the critical behaviour and the
identification of universality classes. We will show that both critical and
complete filling transitions are characterised by universal critical
exponents independent of the intermolecular forces (unless they are
unphysically long-ranged). Interestingly, whilst
the universality of the critical filling transition arises directly as a
consequence of strong fluctuation effects, the universality encountered
at complete filling has a geometrical
and thermodynamic origin which is correctly captured
by mean-field theory. In addition to this, we are able to verify the validity
of the thermodynamic prediction (\ref{one}) and also predict very similar
phenomena occurring when interfaces are pinned in the bulk along a bent line
of weakened bonds.

To begin, we introduce our model. We denote the local height of the liquid-gas
interface relative to the horizontal by $y(x)$ and the height of the wall
itself by $z(x)\equiv\alpha|x|$. The relative local height between the two
is written $\ell(x)\equiv y(x)-z(x)$. The horizontal dimension
of the wall/interface extends over the range $[-X/2,X/2]$ and periodic
boundary conditions are applied at the ends; {\it i.e.}, all
configurations satisfy $y(X/2)=y(-X/2)$ (see Fig.\ (1)). The interface
interacts with the wall via a binding potential arising due to
intermolecular wall-fluid and fluid-fluid forces and is also subject to
thermal fluctuations governed by the stiffness $\Sigma$. We will
concentrate on continuum wall-fluid systems and identify the stiffness
$\Sigma$ with the surface tension. We also set $k_{B}T\!=\!1$ for
convenience. For rather open edges, it is already known from earlier
mean-field studies that the filling transitions are well described
by the simple interfacial model

\begin{equation}
\label{two}
H[y]=\int_{-X/2}^{X/2}\!dx
\;\left[\,{\Sigma\over 2}\left({{dy}\over{dx}}\right)^2
+W(y-z)\,\right]
\end{equation}
which can be justified from analysis of a more general non-linear
model in the small angle $\alpha$ limit \cite{Wedge}.
Before we outline our calculation and
present our main results, we make some preliminary remarks which serve to
establish our notation and also provide some points of comparison.
These concern the {\it planar} limit $\alpha\!=\!0$. For this case,
it is well known that the partition function $Z_{\pi}(\ell_1,\ell_2;X)$
corresponding to the planar fluctuation sum with fixed boundary conditions
$\ell(-X/2)\!=\!\ell_1$ and $\ell(X/2)\!=\!\ell_2$ is given by the
spectral sum (or integral if scattering states are present) \cite{Burkhardt}
\begin{equation}
\label{three}
Z_{\pi}(\ell_1,\ell_2;X)=\sum_{n=0}^{\infty}\;
\psi_{n}^{*}(\ell_1)\,\psi_{n}(\ell_2)\;\;e^{-E_{n}X}
\end{equation}
where the eigenfunctions and eigenvalues satisfy the Schr\"odinger equation
\begin{equation}
\label{four}
-{{1}\over{2\Sigma}}\;\psi_{n}''+W(\ell)\;\psi_{n}=E_{n}\psi_{n}.
\end{equation}
Thus, in the thermodynamic limit $X\rightarrow\infty$, the excess free-energy
is given by $E_0$ which is in turn related to the contact angle (defined only
for $T<T_{w}$ and $\mu=\mu_{sat}(T)$) by $E_0\!=\!-\Sigma\,\Theta_{\pi}^{2}/2$,
valid for small angles $\Theta_{\pi}$ described by the interfacial model.
Similarly, the normalised probability distribution for the interface height
is $P_{\pi}(\ell)\equiv |\psi_{0}(\ell)|^2$. For later purposes, it is also
convenient to define the matrix elements
\begin{equation}
\label{five}
\langle m|f(\ell)|n\rangle\equiv\int \! d\ell\;\;
\psi_{m}^{*}(\ell) f(\ell)\psi_{n}(\ell).
\end{equation}
These will appear in our solutions for the wedge free-energy and
probability distribution function. The binding potentials that we consider
are of the usual form \cite{Dietrich}
\begin{equation}
\label{six}
W(\ell)\,=\,\bar{h}\,\ell+{a\over{\ell^{\,p}}}+{b\over{\ell^{\,q}}};
\hspace{.5cm}\ell>0,
\end{equation}
with $\bar{h}\propto(\mu_{sat}-\mu)$, $a$ and $b$ are effective Hamaker
constants and $q\!>\!p\!>\!0$ allow for general types of intermolecular
potentials. Now, recall that for two dimensional
critical wetting transitions, the critical
behaviour generically belongs to one of three scaling regimes depending
on the values of $p$ and $q$. Specifically, for
$q\!>\!p\!>\!2$, $p\!<\!2$ but $q\!>\!2$ and, finally, $p\!<\!q\!<\!2$,
the behaviour falls into the strong, weak and mean-field fluctuation regimes
with true universality only characteristic of the former \cite{Lipowsky}.
For example, the mean
interface height $\langle\ell\,\rangle_{\pi}\sim(T_{w}-T)^{-\beta_{s}}$
diverges with critical exponents $\beta_{s}\!=\!1$,
$\beta_{s}\!=\!1/(2-p)$, $\beta_{s}\!=\!1/(q-p)$ in the three regimes,
respectively. For complete wetting, corresponding to $\bar{h}\rightarrow 0$
for $T>T_{w}$, there are only the weak and mean-field fluctuation regimes
and the interface height diverges as
$\langle\ell\,\rangle_{\pi}\sim \bar{h}^{-1/3}$ and
$\langle\ell\,\rangle_{\pi}\sim \bar{h}^{-1/(p+1)}$, respectively. These remarks
will serve to illustrate how different the critical behaviour at wedge filling
is to standard planar wetting transitions.

We now turn our attention to the wedge geometry and outline the calculation
of the partition function $Z^{P}_{wedge}(X)$ for the present periodic system.
The advantage of this choice of boundary conditions is that it allows us to
extract the excess wedge free energy $F_{wedge}(\alpha)$ rather easily
from $F_{wedge}^{P}(X)\!=\!-\ln Z^{P}_{wedge}(X)$. To see this, note that
in the thermodynamic limit $X\rightarrow\infty$, the periodic system
reduces to two independent wedges because one must also consider the
contribution from the (inverted) wedge at $x\!=\!\pm X/2$ characterised by
an angle $-\alpha$. First, from $F_{wedge}^{P}(X)$
we subtract the free-energy
$F_{\pi}^{P}(X)$ of a planar system (with periodic boundary conditions)
extending on the same area, {\it i.e.}, a flat wall tilted an angle
$\alpha$ to the horizontal. This can be easily calculated with the same
Hamiltonian, Eq.\ (\ref{two}), but with $z(x)\!=\!\alpha\,x$
(for all $x$ in the range $-X/2\!<\!x\!<\!X/2$).
This defines the excess periodic wedge free-energy:
\begin{equation}
\label{seven}
\Delta F_{wedge}^{P}(X)\equiv F_{wedge}^{P}(X)-F_{\pi}^{P}(X).
\end{equation}
Thus, in the thermodynamic limit, we can write
\begin{equation}
\label{eight}
\lim_{X\rightarrow\infty}\Delta F_{wedge}^{P}(X)=
F_{wedge}(\alpha)-F_{wedge}(-\alpha),
\end{equation}
illustrating the independent contributions from the wedge and inverted
wedge. The partition function $Z^{P}_{wedge}(X)$ is given by the fluctuation
sum over all graphs $y(x)$ or, equivalently, over all relative positions
$\ell(x)\equiv y(x)-z(x)$. Making this change of variable, we can rewrite
the energy, Eq.\ (\ref{two}), of a configuration as
\begin{eqnarray}
\label{nine}
\widetilde{H}[\ell]\;=\;{\Sigma\over 2}\;\alpha^{2}\,X\;+\;
2\;\Sigma\,\alpha\,(\ell_{o}-\ell_{e})\;+\;\\
\int_{-X/2}^{X/2}\!dx
\;\left[\,{\Sigma\over 2}\left({{d\ell}\over{dx}}\right)^2
+W(\ell)\,\right]
\nonumber
\end{eqnarray}
where $\ell_{o}\equiv y(0)$ and $\ell_{e}\equiv y(|X/2|)-z(|X/2|)$ are the
midpoint and edge (relative) interface heights. Consequently,
\begin{eqnarray}
\label{ten}
Z^{P}_{wedge}(X)=e^{-\Sigma\,\alpha^{2}\,X/2}\;\,
\int\!\!\!\int\,d\ell_{o}\,d\ell_{e}\;\;
Z_{\pi}\left(\ell_{e},\ell_{o};{X\over 2}\right)\;\;
\times\\
e^{2\;\Sigma\,\alpha\,(\ell_{e}-\ell_{o})}\;\;
Z_{\pi}\left(\ell_{o},\ell_{e};{X\over 2}\right).
\hspace{2cm}
\nonumber
\end{eqnarray}

Substituting the quantum mechanical result, Eq.\ (\ref{three}), and
taking the thermodynamic limit, we arrive at the general formula for
the wedge free-energy, valid for {\it all} binding potentials,
\begin{equation}
\label{eleven}
F_{wedge}(\alpha)=
-\ln\,\langle\;0\;|\,e^{2\,\Sigma\,\alpha\,\ell}\,|\;0\;\rangle,
\end{equation}
where the inner product is defined in terms of the usual planar system
eigenfunctions (see Eq.\ (\ref{five})). Proceeding in this way, it is also
possible to calculate the probability distribution ${\cal P}(\ell;x)$
for finding the interface at height $\ell$ from the wall at a distance
$x$ along it. We omit details and simply quote our final result obtained
in the thermodynamic limit \cite{Wood},
\begin{eqnarray}
\label{twelve}
{\cal P}(\ell;x)=\!\sum_{n=0}^{\infty}
{{\langle\,n\,|\,e^{2\,\Sigma\,\alpha\,\ell}\,|\,0\,\rangle\;
\psi_{n}^{*}(\ell)\,\psi_{0}(\ell)\;e^{-(E_{n}-E_{0})|x|}}
\over{\langle\;0\;|\,e^{2\,\Sigma\,\alpha\,\ell}\,|\;0\;\rangle}}
\end{eqnarray}
Note that, since the thermodynamic limit is taken first (at finite $x$),
this result pertains to a single wedge system. There is no contribution
due to the inverted wedge at $x\!=\!\pm X/2$. For the probability
distribution of the inverted wedge, one simply reverses the sign of
$\alpha$ and replace
$|x|$ by $|x-X/2|$. Note that when $\alpha\!=\!0$, the wedge free-energy
vanishes and the probability distribution reduces to the standard planar
result ${\cal P}_{\pi}(\ell)$. In addition, for $\alpha\!\neq\!0$,
only the $n\!=\!0$ term survives in the limit $X\rightarrow\infty$
so that ${\cal P}(\ell;x)\rightarrow {\cal P}_{\pi}(\ell)$ infinitely
far from the wedge bottom. The expression for the probability
distribution simplifies considerably if we consider the local height
probability at the mid-point $x\!=\!0$. Writing ${\cal P}(\ell_{0})
\equiv {\cal P}(\ell;0)$, we find
\begin{eqnarray}
\label{thirteen}
{\cal P}(\ell_{0})=
{{|\psi_{0}|^{2}\;\,e^{2\,\Sigma\,\alpha\,\ell_{0}}}
\over{\langle\;0\;|\,e^{2\,\Sigma\,\alpha\,\ell}\,|\;0\;\rangle}},
\end{eqnarray}
which is one of the central results of our paper. We note that both
Eq.\ (\ref{eleven}) and Eq.\ (\ref{twelve}) are consistent since from
the Hamiltonian definition the mean mid-point interface height
satisfy $2\,\Sigma\langle\ell_{0}\rangle=-\partial F_{wedge}(\alpha)/
\partial\alpha$. It is also possible to expand ${\cal P}(\ell;x)$
for small $x$ and derive further explicit results which allow the calculation
of the curvature of various local operators at $x\!=\!0$. In the present
paper, however, we simply concentrate on the properties of the wedge
free-energy $F_{wedge}(\alpha)$ and mid-point height distribution \cite{Wood}.
The essential observation to make here is that, relative to the planar
distribution function ${\cal P}_{\pi}(\ell)$, the mid-point height
probability ${\cal P}(\ell_{0})$ has an exponential boost factor
$e^{2\,\Sigma\,\alpha\,\ell}$ which decreases the pinning effect of
the binding potential (provided $\alpha>0$, of course). Due to this
exponential term, the location (phase boundary) and character of the
filling transition can immediately  traced to the asymptotics of
the planar ground state wave function $\psi_{0}(\ell)$.
If the decay of this function
is too slow, the wedge distribution ${\cal P}(\ell_{0})$ is no longer
defined and the wedge is filled with liquid. Now, for bulk
coexistence ($\bar{h}\!=\!0$) and subwetting temperature ($T\!<\!T_w$),
the asymptotic decay of $\psi_{0}(\ell)$ has the same functional form for
all potentials of the form (\ref{four}):
\begin{equation}
\label{fourteen}
\psi_{0}(\ell)\sim e^{-\Sigma\,\Theta_{\pi}(T)\;\ell};
\hspace{1.5cm}
\bar{h}\!=\!0,\;T\!<\!T_w,\;\ell\rightarrow\infty
\end{equation}
where the specific $p$ and $q$ dependence only enters implicitly
through the temperature dependence of the contact angle
$\Theta_{\pi}(T)\!=\!\sqrt{2|E_{0}|/\Sigma}$. Thus, the location
of the wedge filling transition within the present model exactly matches
with the thermodynamic prediction, Eq.\ (\ref{one}). Moreover, since
$\Theta_{\pi}(T)$ is analytic away from $T_w$, we may identify
$|\Theta_{\pi}(T)-\alpha|\propto T_{\alpha}-T$ and derive the universal critical
singularities for critical filling
\begin{equation}
\label{fifteen}
F_{wedge}(\alpha)\simeq \ln\,(T_{\alpha}-T),
\hspace{1.5cm}
\langle\ell_{0}\,\rangle\sim\;(T_{\alpha}-T)^{-1},
\end{equation}
valid for {\it all} intermolecular forces
(provided $p\!>\!1$). Therefore, the universality
of the critical wedge filling transition far exceeds that encountered
at critical wetting, and is ubiquitous to all realistic solid-fluid interfaces. 
We also note that the transition is fluctuation
dominated since the mid-point roughness
$\xi_{\perp}\equiv
\sqrt{\langle\ell_{0}^{2}\,\rangle-\langle\ell_{0}\,\rangle^2}$
also diverges with the same power law. The universality of the critical
exponents for critical filling is one of the central predictions of
our paper, and can be traced to the large scale interfacial fluctuations
occuring at the bottom of the wedge. In contrast, mean-field
calculations which ignore fluctuations effects, predict highly non-universal
critical behaviour for critical filling. For example, minimization of
the Hamiltonian, Eq.\ (\ref{two}), leads to the prediction that the
mid-point height diverges like
$\langle\ell_{0}\rangle\sim (T_{\alpha}-T)^{-1/p}$ \cite{Wood}. In $d\!=\!2$,
this prediction is only valid for systems with $p\!<\!1$ which do not
correspond to any known physical forces.

Next, we turn our attention to complete filling ocurring
for $\bar{h}\rightarrow0$ and $T\!>\!T_\alpha$. For $\bar{h}\neq 0$, the
asymptotic decay of $\psi_{0}(\ell)$ is faster than exponential due to
the divergent linear term in $W(\ell)$.
This dominates the mid-point probability function ${\cal P}(\ell_0)$
at large distances and ensures that the wedge is only partially filled
when the system is out of bulk two-phase coexistence. The singularity
arising in the evaluation of the wedge free-energy and mean height
$\langle\ell_{0}\rangle$ can be calculated using standard techniques.
For the mean height $\langle\ell_{0}\rangle$, we find that the
leading order behaviour as $\bar{h}\rightarrow 0$ is
\begin{equation}
\label{sixteen}
\langle\ell_{0}\rangle=
\left\{ \begin{array}{cr}
{\displaystyle\Sigma\over\displaystyle 2}\;{\displaystyle{\left(\alpha^{2}-\Theta_{\pi}^{2}(T)\right)}\over{\displaystyle\bar{h}}};
& \hspace{1cm} T_{\alpha}<T\leq T_{w} \\ \\
{{\displaystyle\Sigma\alpha^{2}}\over{\displaystyle 2\bar{h}}} & \hspace{1cm}T>T_{w}
\end{array} \right. 
\end{equation}
which is again independent of the intermolecular potential exponents
$p$ and $q$. In fact, these results are identical
to those derived in the mean-field analysis of K.\ Rejmer {\it et.\ al.\ }
equivalent to a simple minimization of the Hamiltonian (\ref{two}).
Thus, the universality of the complete filling exponents has a geometrical
and thermodynamic origin rather than being a fluctuation related
effect. We mention here that the prediction (\ref{sixteen}) for
the case $T>T_{w}$ is consistent with earlier solid-on-solid model
calculations of complete wetting at a corner \cite{Exact}.
In the light of these transfer-matrix and mean-field results, we
conjecture that the critical singularities occuring at complete filling
are of the form (\ref{sixteen}), independent of the dimensionality.

To end our paper, we mention that very similar phenomena also occur for
interface pinning in the bulk. Recall that an interface is always pinned
along a straight line of weakened bonds, with depinning only occuring
as the strength of the bonds approaches the bulk value \cite{Lipowsky}.
However, this is not the case if the line of weakened bonds has a
bend in it, as can be seen from the present transfer-matrix analysis.
To model this system, we use the same function $z(x)=\alpha|x|$ to describe
the local height of the line of weakened bonds but with a square-well
potential of depth $U$ and range $R/2$,
chosen to mimic the local energy cost due to the bond weakness
(restricting ourself to systems with short-ranged forces).
Whilst in the planar system ($\alpha=0$) the interface has equal
probability of being found above and below the line, the bend at
$x=0$ breaks the symmetry and significantly enhances
the probability of finding the interface above the line of weakened
bonds. In fact, the interface unbinds and depins from the defect line
at a non-zero value of the weakness parameter $U_{\alpha}$ satisfying
$E_{0}(U_{\alpha})\!=\!\Sigma\alpha^{2}/2$ where $E_{0}$ is the
ground state energy of the square-well potential trivially
found from solution of Eq.\ (\ref{four}).

In summary, we have shown through exact transfer-matrix
calculation, that critical filling of liquid in a 2D wedge
is characterized by strongly universal critical behaviour.
Whilst critical filling is
dominated by fluctuations, these do not affect complete filling
which depends solely on the system geometry and whose critical
behaviour is correctly captured by mean-field theory.
Similar behaviour is expected for the 3D wedge. We also point
out similar behaviour occuring at defect lines in the bulk.

A.O.P. is grateful to Prof. S.\ Dietrich for discussions and
hospitality. C.R.\ acknowledges economical support from the E.C.\
under contract ERBFMBICT983229.

\begin{figure}[h]
\label{first}
\vspace*{-1.5cm}
\centerline{\epsfig{file=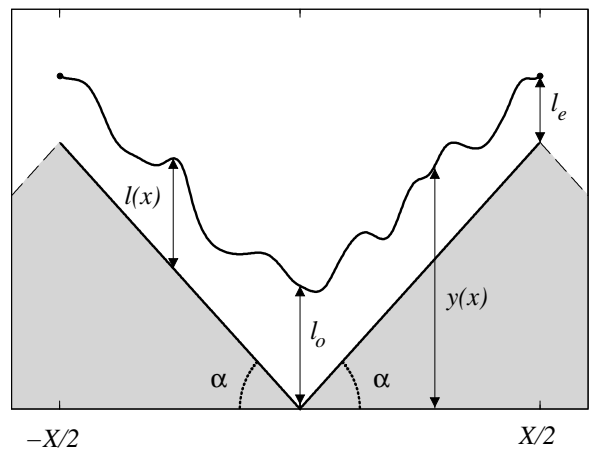,width=12.cm}}
\vspace*{-1.8cm}
\caption{Schematic illustration of an interface configuration $y(x)$ in the
wedge geometry. The function $\ell(x)$ denotes the local distance to the
wall. Other notation is defined in the text.}
\end{figure}

\end{document}